# Hormonal Factors Moderate the Associations Between Vascular Risk Factors and White Matter Hyperintensities

Abdullah Alqarni[1,2,*], Wei Wen[1,3], Ben C.P. Lam[1], John D. Crawford[1], Perminder S. Sachdev[1,3], Jiyang Jiang[1,*]

[1] Centre for Healthy Brain Ageing (CHeBA), Discipline of Psychiatry and Mental Health, Faculty of Medicine, University of New South Wales, Sydney NSW, Australia

[2] Radiology and Medical Imaging Department, College of Applied Medical Sciences, Prince Sattam bin Abdulaziz University, Al-Kharj, Saudi Arabia.

[3] Neuropsychiatric Institute, Prince of Wales Hospital, Randwick, NSW, Australia.

*Corresponding Authors: a.alqarni@student.unsw.edu.au (Abdullah Alqarni); jiyang.jiang@unsw.edu.au (Jiyang Jiang); Centre for Healthy Brain Ageing (CHeBA), Psychiatry and Mental Health, Level 1, AGSM building (G27), Gate 11, Botany Street, UNSW, SYDNEY NSW 2052 AUSTRALIA.

**Abstract**

**Objective:** To examine the moderation effects of hormonal factors on the associations between vascular risk factors and white matter hyperintensities (WMH) in men and women, separately.

**Methods:** WMH were automatically segmented and quantified in the UK Biobank dataset (N = 18,294). Generalised linear models were applied to examine 1) the main effects of vascular (body mass index, hip to waist ratio, pulse wave velocity, hypercholesterolemia, diabetes, hypertension, smoking status) and hormonal (testosterone levels, contraceptive pill, hormone replacement therapy, menopause) factors on WMH, and 2) the moderation effects of hormonal factors on the relationship between vascular risk factors and WMH volumes.

**Results:** In men with testosterone levels one standard deviation (SD) higher than the mean value, increased body mass index and pulse wave velocity, and smoking were associated with higher WMH volumes. The association between body mass index and WMH was more significant in the periventricular white matter regions, whilst the relationship between pulse wave velocity and WMH was restricted to deep white matter regions. Men with low testosterone levels (one SD below the mean level) showed a significant association between hypercholesterolemia and higher deep WMH volumes. Hypertensive women showed higher WMH volumes than women without hypertension regardless of whether hormone replacement therapy was used. However, higher WMH volumes, especially in the deep white matter regions, were found in women who did not use hormone replacement therapy or use it for a shorter duration.

**Conclusion:** These findings highlighted the importance of considering hormonal risk factors in the prevention and management of WMH.

# 1. Introduction

White matter hyperintensities (WMH) are an important imaging biomarker of cerebral small vessel diseases. The aetiology of WMH has not been fully understood but can include ischemia and inflammation [1]. WMH are prominent in the ageing population, and more severe in older adults diagnosed with dementia [2]. WMH appear to be abnormally bright on T2-weighted magnetic resonance imaging (MRI) data. Studies have identified multiple risk factors for WMH, including hypertension [3-5], obesity [6-8], diabetes [9], smoking [3, 10], high homocysteine levels [11], blood lipids abnormality [5, 12], and genetics [13, 14]. Moreover, the effects of these risk factors showed different effects on periventricular and deep white matter regions [6, 13].

Sex differences in WMH burden have been documented in the literature. There has been consistent evidence on women having higher load and faster progression of WMH in ageing [10, 15-18] in comparison with men, although the WMH accumulation in men tend to be more significantly affected by vascular risk factors [6]. Specifically, the impact of vascular risk factors such as hypertension [19, 20] and atherosclerosis [21] were significantly associated with higher WMH in men but not in women. Moreover, obesity was associated with increased WMH in both men and women [7, 8, 22], but greater impacts were found on WMH in men compared to women [6]. On the contrary, the diagnosis of diabetes was associated with higher WMH volumes in women, but not in men [23, 24].

Hormonal factors have been used to investigate the biological basis of sex differences in WMH. Menopause was found to significantly contribute to the development and progression of WMH in some studies [25, 26]. Findings on the effects of hormone replacement therapy (HRT) on the accumulation of WMH burden have been inconsistent, with some studies reporting no effects of HRT on WMH [10, 27, 28], while others showing less WMH burdens in postmenopausal women using HRT [29-31]. In men, available studies on the associations between testosterone levels and WMH did not find significant results [32].

In the current study, we aim to examine whether hormonal factors affect WMH and moderate the associations between vascular risk factors and WMH. Specifically, we examined the effects of testosterone levels, contraceptive pill (CP; history and duration), HRT (history and duration), and menopause (history and post-menopause duration) on WMH, and their moderation effects on the associations between vascular risk factors and WMH volumes in men and women, separately.

## 2. Materials and methods

### 2.1. Study sample

The study sample was drawn from the UK biobank. Brain MRI scans for project ID 37103 were downloaded (N = 23,936). In these 23,936 participants, 5642 were excluded due to failures in WMH segmentation, not passing the visual quality control, missing descriptive data, missing vascular/hormonal risk factors data, or being requested by the participants to be removed from the UK Biobank dataset. Final analyses were performed on 18,294 participants (8618 men, 9676 women) aged between 45 and 80 years (Table 1). The UK biobank was approved by the NHS National Research Ethics Service (ref. 11/ NW/0382), project (10279).

### 2.2. Vascular and hormonal risk factors

Vascular risk factors investigated in this study included body mass index (BMI), hip to waist ratio (HWR), pulse wave velocity (PWV), hypercholesterolemia, diabetes, hypertension (HT), and smoking status. HWR was calculated by dividing waist circumference by hip circumference. The diagnosis of hypercholesterolemia was based on the use of cholesterol lowering medication. Diabetes was defined by combining physician diagnosis and the use of insulin. HT was calculated by combining physician diagnosis and HT medication use. Cigarettes pack per year were used to classify smoking status. In a study of heavy smoking and WMH using UK Biobank data [22], non-smokers (0-10 packs per year), smokers (10-50 packs per year), and heavy smokers (50+ packs per year) were defined. In this study, to ensure a sufficient sample size in each category, we categorised participants into non-smokers (0-10 packs per year) and smokers (10 + packs per year).

The duration of taking CP was calculated by subtracting the age starting using CP from the age of last CP use. HRT duration was calculated in a similar way as CP, where the age of initial HRT was subtracted from the age of last HRT. Post-menopause duration was calculated by subtracting the age at menopause onset from the age at imaging. Testosterone results were extracted from the blood biochemistry analyses in the UK biobank.

## 2.3. MRI acquisition

Brain MRI data acquired at 3 imaging centres (Manchester, Reading, Newcastle) with Siemens Skyra 3T scanners and 32 channel head coils, were used. T1-weighted 3D magnetisation-prepared rapid gradient echo (MPRAGE) and T2-weighted fluid attenuated inversion recovery. data were employed in this study. Details on the scanning parameters have been reported previously [33, 34].

## 2.4. WMH segmentation

To extract and quantify WMH burdens in the UK Biobank, we modified our automated pipeline, UBO detector [35], to accommodate the subtle WMH burdens in UK Biobank participants who are younger than the sample used to develop and validate UBO Detector. Briefly, the warp field from individual T1 space to DARTEL space was first created, and then reversed to warp DARTEL-space atlases (lobar and arterial territories atlases) and masks (lateral ventricles and brain masks) to individual space. After removing non-brain tissue, a k-nearest neighbours classifier was applied to classify candidate clusters generated by FSL FAST into WMH and non-WMH in the native space. WMH voxels with distance from lateral ventricles of less than 12 mm were defined as periventricular WMH (PVWMH). The rest of WMH voxels were categorised as deep WMH (DWMH). The segmentation results were manually checked.

**2.5. Statistical analysis**

Statistical analyses were conducted by using IBM SPSS 27.0 (SPSS, Inc., Chicago, IL). Since the distribution of WMH was non-Gaussian, generalised linear models (GLMs) with Gamma or Tweedie distributions and a log-link function were conducted. Gamma distribution was the main distribution used in the analyses due its better distribution fitness, however, when Gamma distribution did not converge, we used Tweedie distribution. Because of the use of a log-link function, the regression coefficient (b) was exponentiated to aid interpretation (i.e., Exp(b)). For a binary predictor, if Exp(b) is larger than one, then the target group would have an expected percentage higher in WMH of (Exp(b)−1) × 100% than the reference group. If Exp(b) is smaller than one, then the target group would have an expected percentage lower in WMH of (1−Exp(b)) × 100% than the reference group. For a continuous predictor, if Exp(b) is larger than one, then a unit increase in the predictor would be associated with an expected percentage increase in WMH of (Exp(b) −1) × 100%, whereas if Exp(b) is smaller than one, then a unit increase in the predictor would be associated with an expected percentage decrease in WMH of (1−Exp(b)) × 100%.

We first examined the main effects of vascular risk factors on WMH in the whole sample with all vascular risk factors as the independent variables of interest. Demographic variables (age and sex), MRI scanners, and intracranial volumes (ICV) were adjusted for in the models. Moreover, we tested sex moderation effects in the association between each vascular risk factors and WMH in the whole sample while controlling for other vascular risk factors, age, MRI scanners, and ICV.

Since most examined hormonal factors were only present in one sex or intrinsically higher in one sex (e.g., testosterone), their main and moderation effects were tested in men and women separately by applying GLMs on each sub-sample. The main effect of each hormonal factor was tested by adjusting for all vascular risk factors, age, MRI scanners and ICV.

In testing the moderation effect of each hormonal factor on the association between each vascular factor and WMH, an interaction term between the hormonal and vascular risk factor was included, while the variables were mean centred. A hormonal factor moderated the relationship between a vascular risk factor and WMH if there was a significant interaction.

For a significant interaction between a continuous hormonal variable and a vascular risk factor, simple main effects were tested to examine the association between the vascular risk factor and WMH among those who are high in the hormonal factor (one SD above the mean value; +1 SD) and among those who are low (one standard deviation below the mean; -1 SD). The simple main effects were obtained by repeating the analyses which used mean centred hormonal factor variables with those variables centred at 1 SD below the mean, and centred at 1 SD above the mean [36].

For a significant interaction between a binary hormonal variable and a vascular risk factor, simple main effects were tested through coding the corresponding category as 0 and the other category as 1.

Multicollinearity was tested among all independent variables (all Variance Inflation Factors (VIFs) < 2.5). Results with p-values < 0.05 were considered statistically significant. Bonferroni correction was applied to adjust for multiple testing. For our analyses, we examined 3 outcomes (total WMH (TWMH), PVWMH, DWMH) using 2 sets of models (main effect and interaction models), a predictor was considered significant if its p-value was lower than the α level of 0.05/6 = 0.00833.

## 3. Results

### 3.1. The associations of vascular risk factors with WMH in the whole UK Biobank sample and the sex moderation effects on those associations

Women had 44% more PVWMH and 38.4% more DWMH than men ($p < 0.001$). The associations between vascular risk factors and WMH in the whole sample are shown in Table 2. All examined vascular risk factors were significantly associated with TWMH and PVWMH volumes, except that the association between PWV and PVWMH did not remain significant after Bonferroni correction. Hypercholesterolemia, HT, and smoking were also significantly associated with higher DWMH volumes in the whole sample after Bonferroni correction. In the analyses for moderation effects of sex on the associations between vascular risk factors and WMH, sex significantly moderated the relationship of BMI, HWR, hypercholesterolemia, and diabetes on WMH (significant interactions in Supplementary Table. 1). Increased BMI was significantly associated with higher TWMH (Exp(b) = 1.012, $p < .001$, PVWMH (Exp(b) = 1.012, $p < .001$, and DWMH volumes (Exp(b) = 1.011, $p = .004$) in men, but not in women. In men, but not in women, increased HWR was associated with PVWMH (Exp(b) = 1.947, $p < .001$). Men participants with hypercholesterolemia had higher TWMH (Exp(b) = 1.186, $p < .001$), PVWMH (Exp(b) = 1.156, $p < .001$), and DWMH volumes (Exp(b) = 1.277, $p < .001$) compared to women (TWMH; Exp(b) = 1.072, $p < .001$, PVWMH; Exp(b) = 1.074, $p < .001$, DWMH; Exp(b) = 1.060, $p = .04$). Diabetes was significantly associated with PVWMH in women (Exp(b) = 1.146, $p = .004$), but not in men. After the application of Bonferroni correction, only the interaction between sex and BMI (for TWMH, PVWMH, and DWMH) and hypercholesterolemia (for TWMH and DWMH) remained significant.

### 3.2. The associations of vascular risk factors and male-specific hormonal factors with WMH in men

The associations of vascular and hormonal (i.e., testosterone) risk factors with WMH in men are presented in Table 3. While the effects of vascular risk factors in men were comparable to those in the whole sample, testosterone did not show any significant associations with TWMH ($p = .752$), PVWMH ($p = .227$) and DWMH ($p = .070$).

### 3.3. The moderation effects of male-specific hormonal factor on the association between vascular risk factors and WMH in men

The significant moderation effects of the male-specific hormonal factor (i.e., testosterone) are presented in Table 4. The interactions between vascular risk factors and testosterone were significance for BMI, hypercholesterolemia PWV, and smoking. Increased BMI was associated with higher TWMH and PVWMH volumes in men with high testosterone levels (1 SD above mean level; Exp(b) = 1.014, $p < .001$), whereas the relationship between WMH and BMI in men with low testosterone levels (1 SD below mean level) was not significant. Hypercholesterolemia was significantly associated with increased DWMH volumes in men with lower testosterone (Exp(b) = 1.145, $p < .001$), whilst in men with higher testosterone levels there was no significant association between hypercholesterolemia and DWMH. Men with higher testosterone levels showed a significant association between higher PWV and higher DWMH volumes (Exp(b) = 1.021, $p < .001$), but no significant association between PWV and DWMH was found in men with lower testosterone levels. In men with higher testosterone levels, smoking significantly increased TWMH (Exp(b) = 1.278, $p < .001$), PVWMH (Exp(b) = 1.264, $p < .001$), and DWMH (Exp(b) = 1.324, $p < .001$) volumes, but these effects were not significant in men with lower testosterone levels. After Bonferroni correction, only the interaction between smoking and testosterone remained significant.

### 3.4. The associations of vascular risk factors and female-specific hormonal factors with WMH in women

The associations of vascular and hormonal risk factors with WMH in women are presented in Table 5. Hypercholesterolemia, HT, and smoking were associated with higher TWMH, PVWMH and DWMH volumes. Increased HWR was significantly associated with higher TWMH and PVWMH volumes. Moreover, each additional year of CP use was associated with an increase in TWMH and PVWMH by 0.3% (Exp(b) = 1.003, p = 0.017). Interestingly, menopausal women had lower PVWMH volume compared to non-menopausal women (Exp(b) = 0.907, p = .008). Compared to the whole sample, BMI, PWV, and diabetes were not associated with WMH in women. Moreover, testosterone, CP use, post-menopause duration, HRT use, and HRT duration did not have significant associations with WMH in women. The test-wise significant effects of smoking on DWMH, and of HWR and CP duration on TWMH and PVWMH, did not survive Bonferroni correction.

### 3.5. The moderation effect of female-specific hormonal factors on the associations between vascular risk factors and WMH in women.

The significant moderation effects of the continuous hormonal factors on the associations between vascular risk factors and WMH in women are presented in Table 6. The interactions between vascular risk factors and hormonal factors in women were significance for BMI, diabetes, PWV, and HT. In women with lower testosterone, higher BMI was associated with slightly lower DWMH volumes (Exp(b) = 0.991, $p$ = .043). In women with higher testosterone, the diagnosis of diabetes was associated with higher PVWMH volumes (Exp(b) = 1.255, $p$ = .009). In addition, women with shorter HRT duration showed a significant association between higher PWV and higher PVWMH volumes (Exp(b) = 1.010, $p$ = .017). HT increased TWMH and DWMH volumes in women with both higher and lower HRT duration, but the effects were stronger in those with lower HRT duration (Exp(b) for TWMH, 1.352, $p < .001$ vs. 1.244, $p < .001$; Exp(b) for DWMH, 1.499, $p < .001$ vs. 1.296, $p < .001$). Women with lower, but not higher, post-menopause duration showed associations of increased PWV with increased TWMH (Exp(b) = 1.012, $p$ = .017), PVWMH (Exp(b) = 1.013, $p$ = .031) and DWMH (Exp(b) = 1.020, $p$ = .006) volumes. Women with higher post-menopause duration showed negative associations of diabetes with TWMH (Exp(b) = .870, p = .048) and DWMH (Exp(b) = 0.788, p = .013), but those with lower post-menopause duration showed no significant relationship between diabetes and TWMH and a positive relationship with DWMH (Exp(b) = 1.288, $p$ = .039).

The significant moderation effects of the binary hormonal factors on the associations between vascular risk factors and WMH in women are presented in Table 7. Non-menopausal, but not menopausal women, showed associations between higher PWV and higher DWMH volumes (Exp(b) = 1.036, $p$ = .047). HT was associated with higher DWMH volumes in women both with and without HRT, but the effects were stronger in those without HRT (Exp(b) = 1.487, $p < .001$ *vs.* 1.255, $p < .001$). Women with HRT showed a negative association between BMI and DWMH volumes (Exp(b) = 0.991, $p$ = .038).

However, after Bonferroni correction, the moderation effects of HRT duration and binary HRT variable (yes/no) on the associations of HT with DWMH, the moderation effects of post-menopause duration on the association between PWV and TWMH and DWMH, and the moderation effects of post-menopause duration on the association between diabetes and DWMH, remained significant.

## 4. Discussion.

This study examined the main effects of hormonal factors on WMH and the moderation effects of hormonal factors in the association between vascular risk factors and WMH in the UK Biobank dataset (8618 men and 9676 women). The analysis on the moderation effects of sex showed stronger contributions of some vascular risk factors (BMI, Hip-to-waist ratio, and hypercholesterolemia) to the accumulation of WMH in men compared to women, which is consistent with previous studies [6, 10]. In men, testosterone did not show any main effects on WMH. However, testosterone moderated the effects of some vascular risk factors in men, where the effects of BMI, PWV, and smoking on WMH were found at higher, but not lower testosterone levels, and the effects of hypercholesterolemia on WMH were only observed at lower testosterone levels. However, after Bonferroni correction, only the moderating effect of testosterone on smoking remained significant. In women, after Bonferroni correction, the association between HT and DWMH were moderated by HRT use and HRT duration, and the association between diabetes and DWMH was moderated by post-menopause duration. The associations between PWV and TWMH and DWMH were moderated by post-menopause duration and were significant among those experienced shorter post-menopause duration. Diabetes was a significant vascular risk factor for PVWMH in women, however, the association between diabetes and DWMH was moderated by increased post-menopausal duration.

In men, obesity was constantly found to be a major risk factor for WMH, with sex being a significant moderator between obesity measures and increased WMH in men, these findings were consistent with the literature [6, 8]. Despite the lack of direct association between testosterone and WMH in this study and others [32], our results suggest that higher testosterone might have contributed to the effects of obesity on WMH. Moreover, increased PWV indicates arterial stiffness, which was found independently associated with WMH in this study and in the literature [37]. Although the independent association between PWV and DWMH did not survive the Bonferroni correction, at high testosterone levels, the association was significant. Moreover, smoking was one of the major risk factors for WMH in this study and the literature [3, 22]; our results suggest that at higher testosterone levels, smokers might have higher risk of developing all WMH volumes, especially, the high impact on DWMH.

In women, the associations between HRT and WMH have been inconsistent in the literature, where the use of HRT was found to be associated with less WMH in some studies [30, 31], but not others [10, 28]. There were also studies reporting an increase in WMH volume after using HRT [38]. In the current study, HRT was not independently associated with WMH. However, hypertensive women showed significantly more DWMH, especially among those who did not use HRT or used it for shorter time period. At menopause, blood pressure increases gradually, and up to 50% of women are hypertensive at the age of sixty [39]. The use of oral HRT was found to be associated with incident HT [40, 41]. It has been recommended that the intervention with HRT must be initiated early to manage menopause symptoms and benefit from its ability to reduce coronary events [41]. Despite the lack of significance in the association between HRT use duration and WMH in this study, the moderation analysis supports the findings from a longitudinal study on HRT use duration which found that longer-term use could provide neuroprotection for postmenopausal women, especially when initiated early [29]. In addition, further reports support the importance of early initiation of HRT which could provide neuroprotection and warned about late initiation which could be harmful to the brain and not protective [25, 42]. Taken together, despite reports about significant association between HRT and HT, the use of HRT might be beneficial for hypertensive women to add a layer of protection against WMH, especially DWMH.

Menopause marks a significant change for women, where a gradual increase of cardiovascular diseases risk happens due to the hormonal imbalance after menopause, especially estrogen deficiency [39, 43]. The direct association between menopause and WMH has not been found [10]. However, hot flashes (a symptom of menopause) were associated with increased WMH in women with no cardiovascular diseases [44]. Nevertheless, in comparison to peri-menopausal women, cardiovascular risk factors such as blood pressure and lipid profile in post-menopausal women were significantly higher in a large cross-sectional study [45]. In this study, HT and hypercholesterolemia were the major risk factors for WMH in women which support the findings in the literature regarding the effects of these two risk factors on women at older age, given that the majority of women in our sample were post-menopausal. Smoking was another major risk factor for women in our sample. Studies found that the effect of smoking on WMH was significantly higher in women compared to men [10]. Our results showed that non-menopausal women have less PVWMH compared to menopausal women after controlling for age, scanner, ICV, and cardiovascular risk factors. However, the results on menopause might be affected by the number of women with menopause (n = 7458) compared to those without menopause (n = 699), and the age difference between them (mean ± standard deviation of age for menopausal women is 63.78 ± 6.43 (years) and for non-menopausal women 51.35 ± 2.99 (years)). Moreover, in this study, the mean volume of TWMH, PVWMH, and DWMH were higher in menopausal women compared to non-menopausal women. In our analyses of post-menopause duration interaction with cardiovascular risk factors, results were not in line with the literature, where DWMH was higher in participants with shorter post-menopause duration. Therefore, the discrepancies in our study compared to the overall notion in the literature must be interpreted with caution, given the differences in the characteristics between menopausal and non-menopausal women in our sample.

Hormonal factors tend to moderate the effects of vascular risk factors on DWMH, but not PVWMH. DWMH is usually considered to be associated with cerebral amyloid angiopathy (CCA) which affects leptomeningeal and cortical vessels walls by the increased deposition of amyloid-β peptides [46]. Moreover, it has been reported that DWMH might have different genetical phenotypes, different aetiologies, and different associations with different vascular risk factors compared to PVWMH, where higher contribution of cerebral small vessels diseases, pathohistological traits, and axonal loss were more common in DWMH compared to PVWMH [1, 6, 10, 13, 47-49]. Therefore, investigating PVWMH and DWMH separately will provide more insights into the mechanism of WMH development.

There are several strengths and limitations to this study. The WMH volume was extracted using an automated pipeline, which facilitated the investigation of both global and regional WMH. The current study benefited from the large number of participants and the relatively wide age range in UK Biobank data. Nevertheless, the outcomes of our cross-sectional study should be interpreted with caution. Especially, given the UK biobank sample is thought to be generally healthier and have better socioeconomic status compared to the UK population [22]. Another limitation that encourages future investigations is the time difference between the blood biochemistry analysis and the imaging session, which was at least 5 years. Furthermore, the cross-sectional design limits any conclusions on causality. Future longitudinal studies are needed to confirm the findings. This study examined the contribution of sex-specific hormonal factors to the effects of vascular risk factors on WMH. This can be considered as indirect evidence of hormonal factors contributing to the sex differences in the relationship between vascular risk factors and WMH.

## 5. Conclusion.

This study highlights the importance of considering the sex-specific hormonal risk factors and their interaction with vascular risk factors in studying WMH. BMI, hypercholesterolemia, PWV, and smoking contributed differently to WMH in men depending on testosterone levels. In hypertensive women, the use of HRT and longer period of HRT may be beneficial as they moderated the impact of HT on the accumulation of WMH. Management of sex-specific hormonal risk factors and vascular risk factors may substantially contribute to the reduction of WMH burden in men and women.

**Disclosure**

The authors report no disclosures relevant to the manuscript.

**Acknowledgements**

This research was conducted under the UKB Application ID 37103. This research was undertaken with the assistance of resources and services from the National Computational Infrastructure (NCI), which is supported by the Australian Government. Abdullah Alqarni was supported by Prince Sattam Bin Abdulaziz University (Saudi Arabia) and Jiyang Jiang was supported by John Holden Family Foundation.

**References.**

*Table 1. Characteristics of continues and categorical variables in the study sample.*

| Parameter | Sex | N | Mean±Std. / N |
|---|---|---|---|
| TWMH (mm$^3$) | F/M | 9676 / 8618 | 2402.93±3391.61 / 2569.88±3830.95 |
| PVWMH (mm$^3$) | F/M | 9676 / 8618 | 1779.07±2368.81 / 1884.71±2697 |
| DWMH (mm$^3$) | F/M | 9676 / 8618 | 600.14±1208.96 / 657.07±1363.76 |
| Age (years) | F/M | 9676 / 8618 | 62.39±7.27 / 63.74±7.54 |
| ICV (mm$^3$) | F/M | 9635 / 8567 | 1457063.67±131920.44 / 1628214.13±141217 |
| BMI | F/M | 9417 / 8438 | 26.13±4.68 / 27.06±3.9 |
| HWR | F/M | 9458 / 8453 | 0.81±0.06 / 0.9±0.05 |
| PWV (m/s) | F/M | 8945 / 8061 | 9.31±3 / 10.18±2.7 |
| Testosterone (nmol/L) | F/M | 7608 / 7970 | 1.12±0.62 / 12.21±3.53 |
| Post-menopause duration (years) | F | 6926 | 12.16±7.89 |
| Contraceptive pill use duration (years) | F | 7330 | 11.68±8.38 |
| HRT use Duration (years) | F | 9676 | 2.22±6.27 |
| MRI scanner | F,M/ F,M | S1/S3 | F= 8017, M= 7270 / F = 1659, M= 1348 |
| Hypercholesterolemia | F,M/ F,M | No/Yes | F= 8348, M= 6101 / F= 1328, M= 2517 |
| Diabetes | F,M/ F,M | No/Yes | F= 9351, M= 8065 / F= 325, M= 553 |
| Hypertension | F,M/ F,M | No/Yes | F= 8135, M= 6652 / F= 1541, M= 1966 |
| Smoking (Pack per year) | F,M/ F,M | Non-smokers/smokers | F= 8345, M= 6865 / F= 1331, M= 1753 |
| Menopause | F | No/Yes | 699 / 7458 |
| Contraceptive pill use | F | No/Yes | 1331 / 8266 |
| HRT | F | No/Yes | 5992 / 3594 |

TWMH = Total White Matter Hyperintensities; PVWMH = Periventricular White Matter Hyperintensities; DWMH = Deep White Matter Hyperintensities; F= female; M= male; ICV = intracranial volume; BMI = Body Mass Index; HWR = Hip to waist ratio; PWV = Pulse wave velocity; HRT = Hormone replacement therapy; S1 = scanner 1 (Cheadle imaging centre); S3 = scanner 3 (Newcastle imaging centre).

*Table 2. The associations between vascular risk factors and WMH in the whole sample.*

| Risk Factors | TWMH | | PVWMH | | DWMH | |
|---|---|---|---|---|---|---|
| | Exp(b) | P-value | Exp(b) | P-value | Exp(b) | P-value |
| Age | 1.065 | **<.001*** | 1.064 | **<.001*** | 1.069 | **<.001*** |
| Sex | 1.420 | **<.001*** | 1.440 | **<.001*** | 1.384 | **<.001*** |
| BMI | 1.005 | **.004*** | 1.005 | **.003*** | 1.004 | .091 |
| HWR | 1.432 | **.002*** | 1.486 | **.001*** | 1.299 | .103 |
| PWV | 1.006 | **.008*** | 1.006 | **.009** | 1.007 | **.024** |
| Hypercholesterolemia | 1.115 | **<.001*** | 1.105 | **<.001*** | 1.143 | **<.001*** |
| Diabetes | 1.087 | **.008*** | 1.093 | **.004*** | 1.073 | .091 |
| HT | 1.311 | **<.001*** | 1.279 | **<.001*** | 1.415 | **<.001*** |
| Smoking | 1.141 | **<.001*** | 1.142 | **<.001*** | 1.139 | **<.001*** |

Exp(b) is the exponentiated coefficient in a Generalized Linear Model with Gamma or Tweedie distributions and a log-link function. Bold font indicates statistical significance before Bonferroni correction, and p-values with * indicates those survive Bonferroni correction ($\alpha = 0.05/6 = 0.00833$). BMI = Body Mass Index; HWR = Hip-to-waist ratio; HT = hypertension; PWV = pulse wave velocity; TWMH = Total white matter hyperintensities; PVWMH = Periventricular white matter hyperintensities; DWMH = Deep white matter hyperintensities.

*Table 3. The associations between vascular risk factors and WMH in men.*

| Risk Factors | TWMH | | PVWMH | | DWMH | |
|---|---|---|---|---|---|---|
| | Exp(b) | P-value | Exp(b) | P-value | Exp(b) | P-value |
| Age | 1.062 | <.001* | 1.062 | <.001* | 1.063 | <.001* |
| BMI | 1.008 | .009 | 1.009 | .008* | 1.007 | .103 |
| HWR | 1.535 | .042 | 1.672 | .014 | 1.266 | .380 |
| PWV | 1.011 | .003* | 1.011 | .002* | 1.010 | .026 |
| Hypercholesterolemia | 1.073 | .003* | 1.069 | .005* | 1.079 | .011 |
| Diabetes | 1.154 | .001* | 1.174 | <.001* | 1.104 | .063 |
| HT | 1.329 | <.001* | 1.292 | <.001* | 1.442 | <.001* |
| Smoking | 1.164 | <.001* | 1.158 | <.001* | 1.184 | <.001* |
| Testosterone | 1.001 | .752 | 1.003 | .227 | 0.993 | .070 |

Exp(b) is the exponentiated coefficient in a Generalized Linear Model with Gamma or Tweedie distributions and a log-link function. Bold font indicates statistical significance before Bonferroni correction, and p-values with * indicates those survive Bonferroni correction ($\alpha = 0.05/6 = 0.00833$). BMI = Body Mass Index; HWR = Hip-to-waist ratio; HT = hypertension; PWV = pulse wave velocity; TWMH = Total white matter hyperintensities; PVWMH = Periventricular white matter hyperintensities; DWMH = Deep white matter hyperintensities.

*Table 4. The moderation effects of testosterone levels on the associations between vascular risk factors and WMH in men (only significant moderation effects were listed).*

| Vascular risk factor (VRF) | Nature of effect | TWMH | | PVWMH | | DWMH | |
|---|---|---|---|---|---|---|---|
| | | Exp(b) | P-value | Exp(b) | P-value | Exp(b) | P-value |
| **BMI** | Interaction between BMI and testosterone | 1.001 | **.044** | 1.001 | **.044** | - | - |
| | The effect of BMI for high testosterone | 1.014 | **<.001*** | 1.014 | **<.001*** | - | - |
| | The effect of BMI for low testosterone | 1.004 | .248 | 1.005 | .224 | - | - |
| **Hypercholesterolemia** | Interaction between hypercholesterolemia and testosterone | - | - | - | - | 0.981 | **.014** |
| | The effect of hypercholesterolemia for high testosterone | - | - | - | - | 1.002 | .956 |
| | The effect of hypercholesterolemia for low testosterone | - | - | - | - | 1.145 | **<.001*** |
| **PWV** | Interaction between PWV and testosterone | - | - | - | - | 1.003 | **.016** |
| | The effect of PWV for high testosterone | - | - | - | - | 1.021 | **<.001*** |
| | The effect of PWV for low testosterone | - | - | - | - | 0.999 | .842 |
| **Smoking** | Interaction between smoking and testosterone | 1.026 | **<.001*** | 1.024 | **<.001*** | 1.031 | **<.001*** |
| | The effect of smoking for high testosterone | 1.278 | **<.001*** | 1.264 | **<.001*** | 1.324 | **<.001*** |
| | The effect of smoking for low testosterone | 1.065 | .056 | 1.065 | .056 | 1.065 | .132 |

For a significant interaction between a continuous hormonal variable and a vascular risk factor, simple main effects were tested at high level of the hormonal factor (+1SD) and at low level of the factor (-1SD). Exp(b) is the exponentiated coefficient in a Generalized Linear Model with Gamma or Tweedie distributions and a log-link function. Bold font indicates statistical significance before Bonferroni correction, and p-values with * indicates those survive Bonferroni correction ($\alpha = 0.05/6 = 0.00833$). BMI = Body Mass Index; HWR = Hip-to-waist ratio; HT = hypertension; PWV = pulse wave velocity; TWMH = Total white matter hyperintensities; PVWMH = Periventricular white matter hyperintensities; DWMH = Deep white matter hyperintensities.

*Table 5. The associations between vascular risk factors and WMH in women.*

| Risk Factors | TWMH | | PVWMH | | DWMH | |
|---|---|---|---|---|---|---|
| | Exp(b) | P-value | Exp(b) | P-value | Exp(b) | P-value |
| **Age** | 1.067 | **<.001*** | 1.065 | **<.001*** | 1.076 | **<.001*** |
| **BMI** | 1.002 | .455 | 1.002 | .314 | 0.999 | .846 |
| **HWR** | 1.346 | **.042** | 1.347 | **.035** | 1.317 | .177 |
| **PWV** | 1.004 | .175 | 1.004 | .169 | 1.004 | .333 |
| **Hypercholesterolemia** | 1.199 | **<.001*** | 1.179 | **<.001*** | 1.261 | **<.001*** |
| **Diabetes** | 1.005 | .922 | 0.989 | .824 | 1.047 | .502 |
| **HT** | 1.290 | **<.001*** | 1.263 | **<.001*** | 1.380 | **<.001*** |
| **Smoking** | 1.106 | **<.001*** | 1.111 | **<.001*** | 1.095 | **.009** |
| **Testosterone** | 0.999 | .951 | 1.009 | .559 | 0.969 | .146 |
| **CP Use** | 1.024 | .351 | 1.030 | .226 | 1.007 | .832 |
| **CP use duration** | 1.003 | **.017** | 1.003 | **.014** | 1.003 | .077 |
| **Menopause** | 0.936 | .081 | 0.907 | **.008*** | 1.047 | .405 |
| **Menopause duration** | 1.000 | .986 | 1.001 | .608 | 0.996 | .208 |
| **HRT use** | 1.023 | .226 | 1.014 | .450 | 1.051 | .060 |
| **HRT use duration** | 1.002 | .394 | 1.001 | .559 | 1.003 | .233 |

Exp(b) is the exponentiated coefficient in a Generalized Linear Model with Gamma or Tweedie distributions and a log-link function. Bold font indicates statistical significance before Bonferroni correction, and p-values with * indicates those survive Bonferroni correction (α = 0.05/6 = 0.00833). BMI = Body Mass Index; CP = Contraceptive pill; HRT = Hormone replacement therapy; HWR = Hip-to-waist ratio; HT = hypertension; PWV = pulse wave velocity; TWMH = Total white matter hyperintensities; PVWMH = Periventricular white matter hyperintensities; DWMH = Deep white matter hyperintensities.

*Table 6. The moderation effects of continuous female-specific hormonal factors on the relationship between vascular risk factors and WMH in women.*

| Vascular risk factor (VRF) | Nature of effect | TWMH | | PVWMH | | DWMH | |
|---|---|---|---|---|---|---|---|
| | | Exp(b) | P-value | Exp(b) | P-value | Exp(b) | P-value |
| **BMI** | Interaction between BMI and testosterone | 1.007 | **.032** | - | - | 1.010 | **.019** |
| | The effect of BMI for high testosterone | 1.004 | .204 | - | - | 1.004 | .335 |
| | The effect of BMI for low testosterone | 0.995 | .127 | - | - | 0.991 | **.043** |
| **Diabetes** | Interaction between diabetes and testosterone | - | - | 1.285 | **.012** | - | - |
| | The effect of diabetes for high testosterone | - | - | **1.255** | **.009** | - | - |
| | The effect of diabetes for low testosterone | - | - | 0.918 | .279 | - | - |
| **Diabetes** | Interaction between diabetes and post-menopause duration | 0.986 | **.022** | - | - | 0.971 | **.001*** |
| | The effect of diabetes for high post-menopause duration | 0.870 | **.048** | - | - | 0.788 | **.013** |
| | The effect of diabetes for low post-menopause duration | 1.112 | .231 | - | - | 1.288 | **.039** |
| **PWV** | Interaction between PWV and HRT duration | - | - | 0.999 | **.044** | - | - |
| | The effect of PWV for high HRT duration | - | - | 0.998 | .635 | - | - |
| | The effect of PWV for low HRT duration | - | - | 1.010 | **.017** | - | - |
| **PWV** | Interaction between PWV and post-menopause duration | **0.999** | **.008*** | 0.999 | **.032** | 0.998 | **.001*** |
| | The effect of PWV for high post-menopause duration | 0.995 | .241 | 0.997 | .439 | 0.989 | .057 |
| | The effect of PWV for low post-menopause duration | 1.012 | **.017** | 1.013 | **.031** | 1.020 | **.006*** |
| **HT** | Interaction between HT and HRT duration | 0.991 | **.046** | - | - | **0.985** | **.006*** |
| | The effect of HT for high HRT duration | **1.244** | **<.001*** | - | - | **1.298** | **<.001*** |
| | The effect of HT for low HRT duration | **1.352** | **<.001*** | - | - | **1.503** | **<.001*** |

Exp(b) is the exponentiated coefficient in a Generalized Linear Model with Gamma or Tweedie distributions and a log-link function. For a significant interaction between a continuous hormonal variable and a vascular risk factor, simple main effects were tested at high level of the hormonal factor (+ 1SD) and at low level of the factor (-1SD). Bold font indicates statistical significance before Bonferroni correction, and * indicates those survive Bonferroni correction ($\alpha = 0.05/6 = 0.00833$). BMI = Body Mass Index; HRT = Hormone replacement therapy; HT = hypertension; PWV = pulse wave velocity; TWMH =

Total white matter hyperintensities; PVWMH = Periventricular white matter hyperintensities; DWMH = Deep white matter hyperintensities.

*Table 7. The moderation effects of categorical female-specific hormonal factors variables on the relationship between vascular risk factors and WMH in women.*

| Vascular risk factor (VRF) | Nature of effect | TWMH | | PVWMH | | DWMH | |
|---|---|---|---|---|---|---|---|
| | | Exp(b) | P-value | Exp(b) | P-value | Exp(b) | P-value |
| **PWV** | Interaction term between menopause status and PWV | - | - | - | - | 1.037 | .050 |
| | The effect of PWV for menopausal women | - | - | - | - | 0.998 | .887 |
| | The effect of PWV for non-menopausal women | - | - | - | - | 1.036 | **.047** |
| **HT** | Interaction term between HRT and HT | - | - | - | - | 1.185 | **.006*** |
| | The effect of HT for women with HRT | - | - | - | - | **1.255** | **<.001*** |
| | The effect of HT for women without HRT | - | - | - | - | **1.487** | **<.001*** |
| **BMI** | Interaction term between HRT and BMI | - | - | - | - | 1.014 | **.015** |
| | The effect of BMI for women with HRT | - | - | - | - | 0.991 | **.038** |
| | The effect of BMI for women without HRT | - | - | - | - | 1.004 | .274 |

For a significant interaction between a binary hormonal variable and a vascular risk factor, simple main effects were tested through coding the corresponding category as 0 and the other category as 1. Exp(b) is the exponentiated coefficient in a Generalized Linear Model with Gamma or Tweedie distributions and a log-link function. Bold font indicates statistical significance before Bonferroni correction, and p-values with * indicates those survive Bonferroni correction ($\alpha = 0.05/6 = 0.00833$). BMI = Body Mass Index; HRT = Hormone replacement therapy; HT = hypertension; PWV = pulse wave velocity; TWMH = Total white matter hyperintensities; PVWMH = Periventricular white matter hyperintensities; DWMH = Deep white matter hyperintensities.

*Supplementary table 1: The moderation effects of sex on the associations between vascular risk factors and WMH (only significant moderation effects were listed).*

| Vascular risk factor (VRF) | Nature of effect | TWMH | | PVWMH | | DWMH | |
|---|---|---|---|---|---|---|---|
| | | Exp(b) | P-value | Exp(b) | P-value | Exp(b) | P-value |
| **BMI** | Interaction term between BMI and sex | 1.011 | **<.001*** | 1.011 | **<.001*** | 1.012 | **.004*** |
| | The effect of BMI in women | 1.001 | .609 | 1.001 | .537 | 1.000 | .860 |
| | The effect of BMI in men | 1.012 | **<.001*** | 1.012 | **<.001*** | 1.011 | **.001*** |
| **HWR** | Interaction term between HWR and sex | - | - | 1.564 | **.027** | - | - |
| | The effect of HWR in women | - | - | 1.243 | .124 | - | - |
| | The effect of HWR in men | - | - | 1.947 | **<.001*** | - | - |
| **Hypercholesterolemia** | Interaction term between hypercholesterolemia and sex | 0.904 | **.002*** | 0.930 | **.024** | **0.830** | **<.001*** |
| | The effect of hypercholesterolemia in women | 1.072 | **.001*** | 1.074 | **.001*** | 1.060 | **.040** |
| | The effect of hypercholesterolemia in men | 1.186 | **<.001*** | 1.156 | **<.001*** | 1.277 | **<.001*** |
| **Diabetes** | Interaction term between diabetes and sex | - | - | 1.142 | **.030** | - | - |
| | The effect of diabetes in women | - | - | 1.146 | **.004*** | - | - |
| | The effect of diabetes in men | - | - | 1.004 | .942 | - | - |

Exp(b) is the exponentiated coefficient in a Generalized Linear Model with Gamma or Tweedie distributions and a log-link function. Bold font indicates statistical significance before Bonferroni correction, and p-values with * indicates those survive Bonferroni correction ($\alpha = 0.05/6 = 0.00833$). Simple main effects for men and women were derived from analyses using the whole sample (not in men and women separately) in interaction models. BMI = Body Mass Index; HWR = Hip-to-waist ratio; TWMH = Total white matter hyperintensities; PVWMH = Periventricular white matter hyperintensities; DWMH = Deep white matter hyperintensities.